\documentstyle[preprint,aps]{revtex}  
\begin{document}                
\preprint{KNU-TH-30, July 1995}
\draft

\title{Classical Spacetime from Quantum Gravity}
\author{Sang Pyo Kim\footnote{Electronic address:
sangkim@knusun1.kunsan.ac.kr}}
\address{Department of Physics\\
Kunsan National University\\
Kunsan 573-701, Korea}

\maketitle
\begin{abstract}                
We show how classical spacetime emerges from quantum
gravity through the study of
a quantum FRW cosmological model coupled to
a free massive scalar field
using a new asymptotic expansion method of the
Wheeler-DeWitt equation.  It is shown that
the coherent states of
the nonadiabatic basis of a particular
generalized invariant give rise to the quantum
back reaction of matter field proportional
to classical energy and
the Einstein-Hamilton-Jacobi with matter
becomes equivalent to the classical
Einstein equation.
\end{abstract}
\vskip 3cm
\centerline{\it Classical and Qauntum Gravity {\bf 13}, in press}

\newpage

In recent years,
apart from the attempt to endow canonical gravity
with a quantum gravity theory by
overcoming many conceptual and technical
problems, quantum cosmology has also been used
to derive semiclassical gravity in which
one has the quantum field theory in a curved
spacetime and the classical Einstein equation
with the expectation value of
quantum matter field.

If canonical quantum gravity in
the form of the Wheeler-DeWitt (WD)
equation is to have successful and useful
applications in cosmology, there should be
a consistent limiting procedure from a supposed
quantum gravity written formally as
$\hat{{\rm G}}_{\mu\nu} = 8 \pi \kappa \hat{{\rm T}}_{\mu\nu}$,
in which both the gravity and matter field are quantized
at the same time and $\kappa$ is the gravitation constant,
first to the semiclassical gravity written also formally
as ${\rm G}_{\mu\nu} = 8 \pi \kappa
\left< \hat{{\rm T}}_{\mu\nu} \right>$,
in which the spacetime emerges as a classical
background in some sense but the matter field
maintains the status of quantum variables,
and finally down to the classical gravity
${\rm G}_{\mu\nu} = 8 \pi \kappa {\rm T}_{\mu\nu}$,
in which both gravity and matter field are treated as
classical variables.

In the conventional semiclassical approach \cite{Banks},
the WKB approximation of the WD equation
for gravity coupled to a matter field
results in the vacuous Einstein equation in the form
of the Einstein-Hamilton-Jacobi (EHJ) equation
together with a time-dependent functional
Schr\"{o}dinger equation for the matter field.
In order to include quantum back reaction of the matter
field into the EHJ equation
one should keep the expectation value of the quantum matter
field by hand in the Born-Oppenheimer expansion.
However, there are still gaps in this approach to the
semiclassical
gravity \cite{de A}.
Firstly, there is an arbitrariness
in the separation of the WD equation
into the gravitational field equation which should reduce
to the EHJ equation and
the matter field equation which should
describe the Schr\"{o}dinger
equation for the matter field in a curved spacetime
with some gravitational
quantum corrections. Secondly, even after
the cosmological time is introduced from the
EHJ equation and the
Schr\"{o}dinger equation is derived, the form of
quantum states of the matter field should still have
the WKB form to give rise to classical gravity
equated with the matter field in the form of
the EHJ equation.

In this letter we show how a classical spacetime
emerges from a quantum Friedmann-Robertson-Walker
(FRW) cosmological model
coupled to a free massive scalar field in the
semiclassical gravity obtained from a
new asymptotic expansion of the WD
equation \cite{Kim1}.
In the new asymptotic expansion method,
the matter field obeys purely a quantum
equation equivalent to the time-dependent
functional Schr\"{o}dinger equation
with higher order gravitational
quantum corrections, and
semiclassical gravity is described by
the EHJ equation
equated with the quantum back reaction of the scalar field.
In the nonadiabatic basis of a
particular generalized invariant \cite{Cho}, it
is shown that the quantum back reaction of the scalar
field has exactly the same form as the
time-time component of the energy-momentum
tensor in the classical Einstein equation by matching
appropriately quantum numbers to the amplitudes
of classical field.
However, there is a noticeable distinction
between the conventional approach  such as
the WKB approximation or the Born-Oppenheimer
expansion to semiclassical gravity
and the new asymptotic expansion in that
in the former one should use the WKB form
of wave function at some stage of its
derivation to obtain the classical Einstein
equation, whereas in the latter one keeps
the quantum status without
assuming the WKB wave function for the
scalar field and the quantum back reaction
just yields the time-time component of the
energy-momentum tensor.

The FRW cosmology has a homogeneous and
isotropic spacetime manifold with the metric
\begin{equation}
ds^2 = - N^2(t) dt^2 + R^2 (t) d\Omega_3^2,
\end{equation}
where $N$ is the lapse function and $R(t)$
is the scale factor depending only on $t$.
The time will be scaled in unit of $c = 1$, and
the Planck mass squared will thus denote $m_P^2
= \frac{\hbar}{8 \pi \kappa}$.
The action for the FRW cosmology coupled to a
free massive scalar field, which
is also homogeneous and isotropic, i.e.,
depends only on time $t$, takes the form
\begin{equation}
S = \int dt \left[ - \frac{m_P^2}{\hbar} R^3
\left(\frac{1}{2N} \left(\frac{\dot{R}}{R} \right)^2
+ N \left(-\frac{k}{2R^2}
+ \frac{\Lambda}{6} \right) \right)
+ R^3 \left(\frac{1}{2N} \dot{\phi}^2 -
\frac{Nm^2}{2} \phi^2 \right) \right].
\end{equation}
The classical equations of motion are obtained
by taking variation with respect to $N$:
\begin{equation}
- \frac{m_P^2}{\hbar} \left( \frac{R\dot{R}^2}{2}
+ \frac{kR}{2}
- \frac{\Lambda R^3}{6} \right)
+ R^3 \left(\frac{\dot{\phi}^2}{2} +
\frac{m^2}{2} \phi^2 \right) = 0,
\label{cl Ein eq}
\end{equation}
by taking variation with respect to $R$:
\begin{equation}
\frac{m_P^2}{\hbar} \left( \frac{d}{dt}(R\dot{R})
-\frac{\dot{R}^2}{2}
+ \frac{k}{2}
- \frac{\Lambda R^2}{2} R^2 \right)
+ 3 R^2 \left(\frac{\dot{\phi}^2}{2} -
\frac{m^2}{2} \phi^2 \right) = 0,
\end{equation}
and by taking variation with respect to $\phi$:
\begin{equation}
\frac{d}{dt} \left(R^3 \dot{\phi} \right)
+ m^2 R^3 \phi = 0.
\label{cl eq}
\end{equation}
We rewrite the classical equation
(\ref{cl Ein eq}) in the form
\begin{equation}
\left( \frac{\dot{R}}{R} \right)^2
+ \frac{k}{R^2} - \frac{\Lambda}{3}
= 16 \pi \kappa \frac{1}{R^3} {\rm T}_{00}
\end{equation}
where ${\rm T}_{00}$ is the time-time component of
the energy-momentum stress tensor
\begin{equation}
{\rm T}_{00} =
 \frac{1}{2} R^3 \dot{\phi}^2 +
\frac{1}{2}m^2 R^3 \phi^2.
\end{equation}

Up to some operator ordering ambiguity, we quantize
the Hamiltonian a la Dirac quantization to
obtain the WD equation
\begin{equation}
\left[ \frac{\hbar^3}{2m_P^2 R}
\frac{\partial^2}{\partial R^2} + \frac{m_P^2}{\hbar}
\left(- \frac{k R}{2} + \frac{\Lambda R^3}{6} \right)
- \frac{\hbar^2}{2R^3} \frac{\partial^2}{\partial \phi^2}
+ \frac{m^2 R^3}{2} \phi^2
 \right] \Psi (R, \phi) = 0.
\end{equation}
The key tool of the new asymptotic expansion method
is to search for the wave function of the form
\begin{equation}
\Psi (R, \phi) = \psi (R) \Phi (\phi, R).
\label{w f}
\end{equation}
Here $\psi$ and $\Phi$ are still quantum states of gravity
and the scalar field, respectively.
The classical spacetime will emerge later only
after finding the wave function for gravity
in the WKB form with the quantum
back reaction of scalar field.
The wave function (\ref{w f}) has the form obtained
from the adiabatic expansion or superadiabatic expansion of the WD
equation \cite{Kim2}.
Any quantum state of the scalar field can be expanded
by the basis of some Hermitian operator
relevant to the matter field Hamiltonian:
\begin{equation}
\Phi (\phi, R) = \sum_k c_k (R) \left|
\Phi_k (\phi, R) \right>.
\end{equation}
The main result is the separation of the WD
equation into the gravitational field equation
\begin{equation}
\left[\frac{\hbar^3}{2m^2_P R}
\frac{\partial^2}{\partial R^2}
+ \frac{m^2_P}{\hbar} \left(- \frac{kR}{2} +
\frac{\Lambda R^3}{6} \right)
+ :{\rm H}:_{nn} (R) \right] \psi(R) = 0,
\end{equation}
and in the limit $\frac{\hbar^2}{m_P^2}
\rightarrow 0$ the asymptotic matter field equation
\begin{equation}
i \hbar \frac{\partial}{\partial t} c_n
+ \left(\Omega_{nn}^{(1)} - H_{nn} + :{\rm H}:_{nn} \right) c_n
+ \sum_{k \neq n} \left(\Omega_{nk}^{(1)} - {\rm H}_{nk} \right)
c_k  = 0,
\end{equation}
where
\begin{eqnarray}
\Omega_{nk}^{(1)} &=& i \hbar \left<\Phi_n (\phi, R)
| \frac{\partial}{\partial t} |
\Phi_k (\phi, R) \right>,
\nonumber\\
{\rm H}_{nk} &=& \left<\Phi_n (\phi, R)
| \hat{{\rm H}}_m |
\Phi_k (\phi, R) \right>,
\end{eqnarray}
and the expectation value of normal ordered Hamiltonian
\begin{equation}
:{\rm H}:_{nn} = \left<\Phi_n (\phi, R)
| :\hat{{\rm H}}_m :|
\Phi_n (\phi, R) \right>,
\end{equation}
through the introduction of cosmological time
\begin{equation}
\frac{\partial}{\partial t} = {\rm Im} \left(
\frac{\hbar^2}{m_P^2} \frac{1}{R \psi}
\frac{\partial \psi}{\partial R}
\frac{\partial}{\partial R} \right).
\end{equation}
For the gravitational wave function in the WKB form
\begin{equation}
\psi(R) = f(R) \exp \left[i \frac{m_P^2}{\hbar^2}
S(R) \right]
\end{equation}
the cosmological time becomes
\begin{equation}
\frac{\partial}{\partial t} =
\frac{1}{R} \frac{\partial S(R)}{\partial R}
\frac{\partial}{\partial R}.
\end{equation}
From the gravitational wave equation we obtain
the EHJ equation
\begin{equation}
\frac{1}{2R} \left(\frac{\partial S}{\partial R} \right)^2
+ \frac{kR}{2} - \frac{\Lambda R^3}{6}
= 8 \pi \kappa :{\rm H}:_{nn}(R),
\label{EHJ eq}
\end{equation}
which can be rewritten as
\begin{equation}
\left( \frac{\dot{R}}{R} \right)^2
+ \frac{k}{R^2} - \frac{\Lambda}{3}
= 8 \pi \kappa \frac{1}{R^3} :{\rm H}:_{nn}(R).
\end{equation}

On the other hand, the matter field Hamiltonian is a
time-dependent harmonic oscillator of the form
\begin{equation}
\hat{{\rm H}}_m = {\rm T}_{00} = \frac{1}{2R^3} \pi_\phi^2 +
\frac{1}{2}m^2 R^3 \phi^2.
\end{equation}
In the nonadiabatic bases
\begin{equation}
\hat{I} (\pi_\phi, \phi, R) \left|\Phi_n (\phi,R)
\right>
=
\lambda_n  \left|\Phi_n (\phi,R) \right>.
\end{equation}
of the generalized invariants, which obey the invariant
equation
\begin{equation}
\frac{\partial}{\partial t} \hat{I}
- \frac{i}{\hbar} \left[\hat{I},\hat{{\rm H}}_m \right]
= 0,
\end{equation}
there is a well-known decoupling theorem \cite{Lewis}
\begin{equation}
{\rm H}_{nk} (R) = \Omega_{nk}^{(1)}
\end{equation}
for $n \neq k$. Then the asymptotic matter field equation
becomes a diagonal equation whose solution is
given by
\begin{equation}
c_n (t) = c_n (t_0) \exp \left[\frac{i}{\hbar} \int
\left( \Omega_{nn}^{(1)}
- {\rm H}_{nn} + :{\rm H}:_{nn} \right) dt \right].
\label{as s}
\end{equation}
It should be noted that the asymptotic quantum
state is the exact quantum state
of the Schr\"{o}dinger equation
\begin{equation}
i \hbar \frac{\partial}{\partial t}
\Phi (\phi, R) = \hat{{\rm H}}_m (\pi_\phi, \phi, R)
\Phi (\phi, R)
\end{equation}
except for the time-dependent phase factor
$\exp \left[\frac{i}{\hbar} \int
:{\rm H}:_{nn} dt \right]$.

First we find the particular
second order generalized invariant
of the form \cite{Cho}
\begin{equation}
\hat{I} = \frac{1}{2} \left(\hat{I}_+ \hat{I}_-
+ \hat{I}_- \hat{I}_+ \right),
\end{equation}
where
\begin{eqnarray}
\hat{I}_+ &=& \phi^* (t) \hat{\pi}_\phi
- R^3 (t) \dot{\phi}^* (t) \hat{\phi},
\nonumber\\
\hat{I}_- &=& \phi (t) \hat{\pi}_\phi
- R^3 (t) \dot{\phi} (t) \hat{\phi},
\end{eqnarray}
in terms of one of classical solutions of
Eq. (\ref{cl eq}) such that
\begin{eqnarray}
&& R^3 (t) \left(\phi^*(t) \dot{\phi}(t)
- \phi(t) \dot{\phi}^*(t) \right) = i,
\nonumber\\
&& {\rm Im} \left(\frac{\dot{\phi}(t)}{\phi(t)} \right)
< 0.
\label{con}
\end{eqnarray}
Then $\hat{I}_+$ acts as the creation operator
$\hat{A}^\dagger (t)$ and $\hat{I}_-$
as the annihilation operator $\hat{A} (t)$.
The ground state quantum state is  given by
\begin{equation}
\left<\phi|\Phi_0(\phi,R) \right>
= \frac{1}{(2\pi \hbar |\phi(t)|^2 )^{1/4}}
\exp\left[i \frac{R^3 \dot{\phi}(t) }{2\hbar \phi(t)}
\phi^2 \right],
\end{equation}
and the $n$th quantum state by
\begin{equation}
\left<\phi|\Phi_n(\phi,R) \right>
= \frac{1}{(2\pi \hbar |\phi(t)|^2 )^{1/4}}
\frac{1}{\sqrt{2^n n!}}
\left(i \frac{\phi^*(t)}{|\phi(t)| } \right)^n
H_n \left( \frac{\phi}{\sqrt{2\hbar} |\phi(t)|} \right)
\exp\left[i \frac{R^3 \dot{\phi}(t) }{2\hbar \phi(t)}
\phi^2 \right],
\end{equation}
where $H_n$ is the $n$th Hermite polynomial.
From the quantum back reaction of the scalar field
\begin{equation}
:{\rm H}:_{nn} (t) = n \hbar R^3(t) \left(
\dot{\phi}(t) \dot{\phi}^*(t) + m^2 \phi (t)
\phi^*(t)
\right)
\end{equation}
we obtain the EHJ equation
with the quantum back reaction in the nonadiabatic basis
\begin{equation}
\left( \frac{\dot{R}}{R} \right)^2
+ \frac{k}{R^2} - \frac{\Lambda}{3}
= 8 \pi n \hbar \kappa  \left(
\dot{\phi}(t) \dot{\phi}^*(t) + m^2 \phi (t)
\phi^* (t)
\right).
\end{equation}
The ground state ($n = 0$) of the scalar field
leads to the Einstein vacuum equation. Furthermore,
one can show that the semiclassical gravity
reduces classical gravity
by identifying the amplitude of classical field
$\phi_c = \phi_0 \phi_q$ where $\phi_0 =
\sqrt{n\hbar}$ and  $\phi_q$
satisfies the condition (\ref{con}). In particular,
the classical field energy is proportional to
the field squared and the quantum energy to $n \hbar$,
and therefore for a large quantum number $n$ one may expect
the correspondence
$\phi_0 = \sqrt{n\hbar}$.

To summarize, we showed how classical spacetime obeying
the Einstein-Hamilton-Jacobi equation with the back
reaction of classical matter field, which is
equivalent
to classical Einstein equation equated with
the matter field, can emerge through the investigation
of the quantum FRW cosmological model minimally
coupled to a free massive scalar field.
It differs somewhat conceptually and technically
from the conventional approach. The quantum status of
matter field has been still kept through the asymptotic
limiting procedure, a Born-Oppenheimer expansion,
for the gravitational field equation. There
needs no restriction on the form of the wave function
of matter field as far as the cosmological time
is appropriately defined.
The remaining open question is to see how
the method used in this letter can be applied
successfully to the quantum FRW cosmological model
minimally coupled to a scalar field with an arbitrary
potential and whether classical spacetime emerges
really from quantum gravity.
Even though to find the generalized invariant is not
so simple beyond a quadratic potential, the extension
to a general potential is worthy to be worked out.

\acknowledgments

This work was supported in part by the
Non-Directed Research Fund, Korea Research Foundation, 1994.


\begin{references}

\bibitem{Banks} Lapchinsky V and Rubakov V A 1979 {\it Acta Phys.
Pol.} B {\bf 10} 1041\\
Banks T 1985 {\it  Nucl. Phys.} {\bf B249} 332 \\
Brout R 1987
{\it Found. Phys. Phys.} {\bf 17} 603\\
Brout R and  Weil D 1987 {\it Phys. Lett.}
 B {\bf 192} 318\\
Brout R 1987
{\it Z. Phys.} B {\bf 68} 339\\
Kiefer C 1987 {\it Class. Quantum Grav.} {\bf 4} 1369\\
Singh T P and Padmanabhan T 1989
{\it Ann. Phys.} {\bf 196} 296\\
Brout R and Venturi G 1989
{\it Phys. Rev.} D {\bf 39} 2436\\
Kowalski-Glikman J and Vink J C 1990
{\it Class. Quantum Grav.} {\bf 7} 901\\
Singh T P 1990
{\it Class. Quantum Grav.} {\bf 7} L149\\
Balbinot R, Barletta A and Venturi G 1990
{\it Phys. Rev.} D
{\bf 41} 1848\\
Paz J P and Sinha S 1991
{\it Phys. Rev.} D {\bf 44} 1038\\
Kiefer C and T. P. Singh T P 1991 {\it Phys. Rev.} D {\bf 44} 1067\\
Vink J C 1992 {\it Nucl. Phys.} B {\bf 369} 707\\
Gundlach C 1993 {\it Phys. Rev.} D {\bf 48} 1700\\
Datta D P 1993 {\it Mod. Phys. Lett.} A {\bf 8} 191; 2523(E)\\
Datta D P 1993 {\it Mod. Phys. Lett.} A {\bf 8} 601\\
Datta D P 1993 {\it Phys. Rev.} D {\bf 48} 5746
\bibitem{de A} de Alwis S P and MacIntire 1994
{\it Phys. Rev.} D {\bf 50} 5164
\bibitem{Kim1} Kim S P 1995 {\it Phys. Rev.} D {\bf 52}
3382\\
Kim S P 1995 {\it Phys. Lett.} A {\bf 205} 359
\bibitem{Cho} Cho K H, Kim S P and Soh D S 1995
{\it Functional Schr\"{o}dinger Ficture Field Theory
in Curved Spacetime : Free Scalar Field} KNU-TH-29
\bibitem{Lewis} Lewis H R Jr. and  Riesenfeld W B 1969
{\it J. Math. Phys.} {\bf 10} 1458
\bibitem{Kim2} Kiefer C 1988 {\it Phys. Rev.} D {\bf 38} 1761\\
Kim S P 1992 {\it Phys. Rev.} D {\bf 46} 3403\\
Kim S P, Kim J, and Soh K S 1993
{\it Nucl. Phys.} B {\bf 406} 481

\end{references}
\end{document}